\documentclass[twoside,twocolumn,9pt]{article}
\usepackage{extsizes}
\usepackage[super,sort&compress,comma]{natbib} 
\usepackage[version=3]{mhchem}
\usepackage[left=1.5cm, right=1.5cm, top=1.785cm, bottom=2.0cm]{geometry}
\usepackage{balance}
\usepackage{mathptmx}
\usepackage{sectsty}
\usepackage{graphicx} 
\usepackage{lastpage}
\usepackage[format=plain,justification=justified,singlelinecheck=false,font={stretch=1.125,small,sf},labelfont=bf,labelsep=space]{caption}
\usepackage{float}
\usepackage{fancyhdr}
\usepackage{fnpos}
\usepackage[english]{babel}
\addto{\captionsenglish}{%
  
}
\usepackage{array}
\usepackage{droidsans}
\usepackage{charter}
\usepackage[T1]{fontenc}
\usepackage[usenames,dvipsnames]{xcolor}
\usepackage{setspace}
\usepackage[compact]{titlesec}
\usepackage{hyperref}

\usepackage{epstopdf}

\definecolor{cream}{RGB}{222,217,201}

\begin{document}

\pagestyle{fancy}
\thispagestyle{plain}
\fancypagestyle{plain}{
\renewcommand{\headrulewidth}{0pt}
}

\makeFNbottom
\makeatletter
\renewcommand\LARGE{\@setfontsize\LARGE{15pt}{17}}
\renewcommand\Large{\@setfontsize\Large{12pt}{14}}
\renewcommand\large{\@setfontsize\large{10pt}{12}}
\renewcommand\footnotesize{\@setfontsize\footnotesize{7pt}{10}}
\makeatother

\renewcommand{\thefootnote}{\fnsymbol{footnote}}
\renewcommand\footnoterule{\vspace*{1pt}%
\color{cream}\hrule width 3.5in height 0.4pt \color{black}\vspace*{5pt}} 
\setcounter{secnumdepth}{5}

\makeatletter 
\renewcommand\@biblabel[1]{#1}            
\renewcommand\@makefntext[1]%
{\noindent\makebox[0pt][r]{\@thefnmark\,}#1}
\makeatother 
\renewcommand{\figurename}{\small{Fig.}~}
\sectionfont{\sffamily\Large}
\subsectionfont{\normalsize}
\subsubsectionfont{\bf}
\setstretch{1.125} 
\setlength{\skip\footins}{0.8cm}
\setlength{\footnotesep}{0.25cm}
\setlength{\jot}{10pt}
\titlespacing*{\section}{0pt}{4pt}{4pt}
\titlespacing*{\subsection}{0pt}{15pt}{1pt}

\fancyfoot{}
\fancyfoot[LO,RE]{\vspace{-7.1pt}\includegraphics[height=9pt]{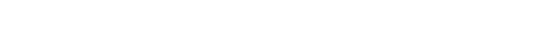}}
\fancyfoot[CO]{\vspace{-7.1pt}\hspace{13.2cm}\includegraphics{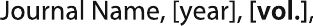}}
\fancyfoot[CE]{\vspace{-7.2pt}\hspace{-14.2cm}\includegraphics{RF}}
\fancyfoot[RO]{\footnotesize{\sffamily{1--\pageref{LastPage} ~\textbar  \hspace{2pt}\thepage}}}
\fancyfoot[LE]{\footnotesize{\sffamily{\thepage~\textbar\hspace{3.45cm} 1--\pageref{LastPage}}}}
\fancyhead{}
\renewcommand{\headrulewidth}{0pt} 
\renewcommand{\footrulewidth}{0pt}
\setlength{\arrayrulewidth}{1pt}
\setlength{\columnsep}{6.5mm}
\setlength\bibsep{1pt}

\makeatletter 
\newlength{\figrulesep} 
\setlength{\figrulesep}{0.5\textfloatsep} 

\newcommand{\topfigrule}{\vspace*{-1pt}%
\noindent{\color{cream}\rule[-\figrulesep]{\columnwidth}{1.5pt}} }

\newcommand{\botfigrule}{\vspace*{-2pt}%
\noindent{\color{cream}\rule[\figrulesep]{\columnwidth}{1.5pt}} }

\newcommand{\dblfigrule}{\vspace*{-1pt}%
\noindent{\color{cream}\rule[-\figrulesep]{\textwidth}{1.5pt}} }

\makeatother

\twocolumn[
  \begin{@twocolumnfalse}
{\includegraphics[height=30pt]{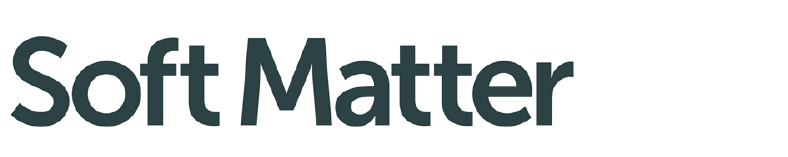}\hfill\raisebox{0pt}[0pt][0pt]{\includegraphics[height=55pt]{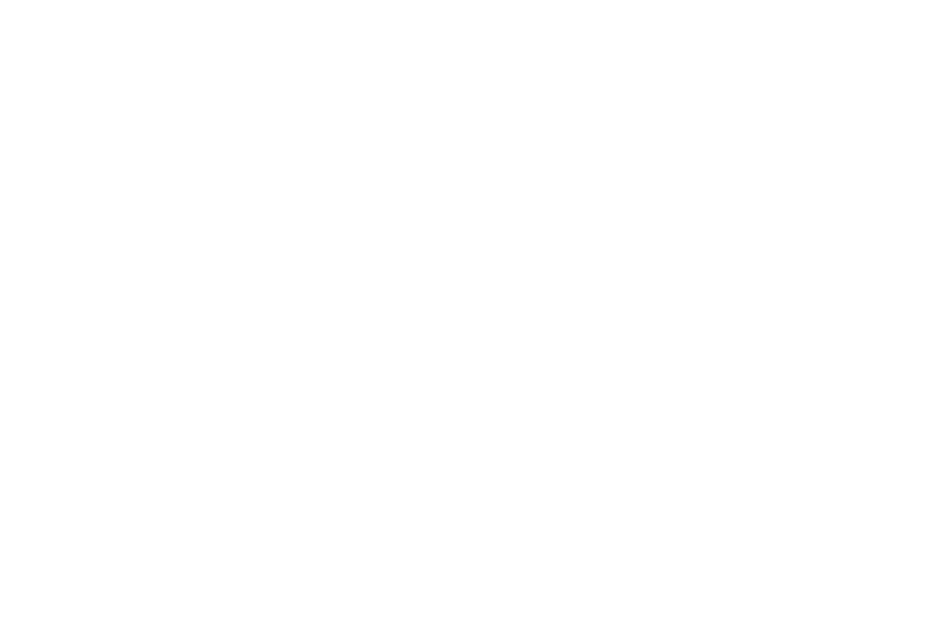}}\\[1ex]
\includegraphics[width=18.5cm]{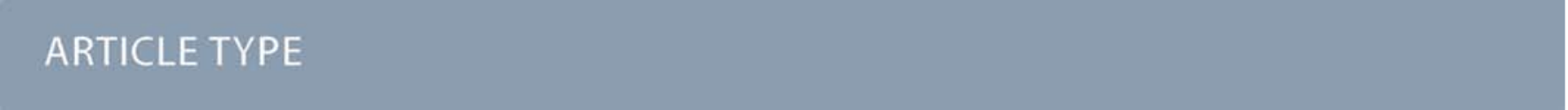}}\par
\vspace{1em}
\sffamily
\begin{tabular}{m{4.5cm} p{13.5cm} }

\includegraphics{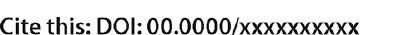} & \noindent\LARGE{\textbf{The Coil-Globule transition in self-avoiding active polymers
}} \\
\vspace{0.3cm} & \vspace{0.3cm} \\

 & \noindent\large{S. Das,\textit{$^{a,b}$} N. Kennedy,\textit{$^{a}$} and A. Cacciuto$^{\ast}$\textit{$^{a}$}} \\

\includegraphics{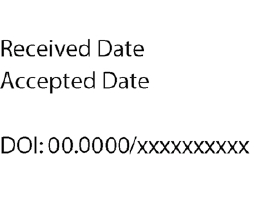} & \noindent\normalsize{We perform numerical simulations of an active fully flexible self-avoiding polymer as a function of the quality of the embedding solvent described in terms of an effective monomer-monomer interaction. Specifically, by extracting the Flory exponent of the active polymer under different conditions, we are able to pin down the location of the coil-globule transition for different strength of the active forces. Remarkably, we find that a simple rescaling of the temperature is capable of qualitatively capture the dependence of the $\Theta$-point of the polymer with the amplitude of the active fluctuations. We discuss the limits of this mapping, and suggest that a negative  active pressure between the monomers,  not unlike the one that has already been found in suspensions of active hard spheres, may also be present in active polymers. } \\

\end{tabular}

 \end{@twocolumnfalse} \vspace{0.6cm}

  ]

\renewcommand*\rmdefault{bch}\normalfont\upshape
\rmfamily
\section*{}
\vspace{-1cm}


\footnotetext{\textit{$^{a}$~Department of Chemistry, Columbia University, 3000 Broadway, New York, NY 10027}}
\footnotetext{\textit{$^{b}$~Department of Polymer Science and Engineering, University of Massachusetts, 120 Governors Drive, Amherst, MA 01003}}
\footnotetext{$^{\ast}$~ac2822@columbia.edu}




\section{Introduction}
Many of the important dynamical properties of biological polymers such as actin filaments and microtubules are due to their out-of-equilibrium behavior. Using ATP (Adenosine triphosphate) or GTP (Guanosine triphosphate) these filaments are capable of powering their locomotion and exert significant forces against the cell membrane.
This idea of generating local forces by capitalizing on available energy sources in the environment is at the forefront of the research on active materials, and includes the study of all those biological~\cite{wang_one_2015} or synthetic systems containing components that can be driven far-from-equilibrium by exploiting local chemical, electrical or thermal gradients~\cite{dey_chemically_2017}. 

Among the many varieties of synthetic active system, the study of active polymers has seen a recent 
burst in theoretical activities~\cite{Loi2011Oct,Kaiser2014Jul,Harder2014Dec,Ghosh2014Sep,Shin2015Oct,Kaiser2015Mar,Samanta2016Apr,Eisenstecken2016Aug,Chelakkot2014Mar,Isele-Holder2015Sep,Isele-Holder2016Oct,Kaiser2015Mar,Nogucci2016May,Bianco2018Nov,Harder2018Feb} 
(see also \cite{Winkler2017Aug} for a brief review on the subject and references therein).
This is because active polymers can be considered as a minimal model where the competition between thermodynamic and active forces can be systematically studied in a system that can also undergo conformational transformations.

Broadly speaking, two models for active polymers have been put forward. 
In the first model activity is introduced as a local force that is tangential to the backbone of the polymer at every point, and it is appropriate to describe the collective behavior of suspensions of active filaments interconnected by molecular motors~\cite{Schaller2010Sep,Schaller2011Mar}) or strings of Janus dipolar particles~\cite{Yan2016Jul}. In the second model, developed to mimic the behavior of a passive filament undergoing random active fluctuations in an embedding active fluid, the direction of the active forces applied to each monomer is subject to random brownian rotations. 

For this second active Brownian polymer model, a number of studies have measured the radius of gyration, $R_{\rm g}$, of the polymer as a function of the strength of the active forces, and found  a non-monotonic dependence on the strength of the active forces for self-avoiding chains.
The polymer undergoes a small compression for intermediate activities and re-expand when the active forces become sufficiently large; yet, for a fixed active force, $R_{\rm g}$ is expected to follow the Flory scaling law exhibited by its passive counterpart~\cite{Kaiser2015Mar,Bianco2018Nov,deviations2019,eisenstecken2017conformational,anand2020conformation} when plotted as a function of $N$ (at least for activities that are not too large). 
This non-trivial result suggests that some of the statistical methods used to study passive polymers could be also used to understand the behavior of active polymers. We have recently shown that such a mapping can only be applied under specific circumstances when dealing with active polymers under confinement~\cite{deviations2019}.
In this paper, we explore another important property of polymers, one that is at the core of their biological function and phase behavior: the coil-to-globule transition, that takes place when a flexible polymer is in the presence of a bad solvent\cite{Gennes1979Nov,Khokhlov2002Mar}.
The radius of gyration of a flexible self-avoiding polymer in a good solvent scales with its size, $N$, as $R_g\sim N^{\nu}$ with $\nu\simeq {3/(2+d)}$. Here $d$ is the dimensionality of the embedding space. When the polymer is immersed in a bad solvent, it collapses into a globule and its radius of gyration scales as $R_g\sim N^{1/d}$. When the repulsion between the monomers exactly matches their effective attraction generated by solvent avoidance, +at the polymer $\Theta$ temperature, the polymer behaves ideally and follows the universal scaling law $R_g\sim N^{1/2}$. In this paper we show that the same transition holds for active polymers, and that a simple theoretical framework can be used to understand how the active forces affect the behavior of the polymer undergoing this transition.

\section{Model}
Our model for a three dimensional flexible, self-avoiding active polymer consists of $N$ monomers of diameter $\sigma$ linearly connected with harmonic bonds. Each monomer of the polymer chain undergoes Brownian motion in three dimensions at a temperature $T$ and is subject to an active velocity of magnitude $v_p$ according to the following translational and rotational equations of motion: 
\begin{equation}
\label{eq:trans} 
\frac{d\pmb{r}(t)}{dt} = \frac{1}{\gamma} \pmb{f}(t) +   v_p \, \pmb{\hat{q}}(t) + \sqrt{2D}\,\pmb{\xi}(t),
\end{equation}

\begin{equation}
\label{eq:rot}
\frac{d \pmb{{\hat{q}}}(t) }{dt} = \sqrt{2D_r}\, \pmb{\xi}_r(t) \times \pmb{\hat{q}}(t).
\end{equation}
Active forcing is directed along a predefined orientation unit vector $\pmb{\hat{q}}$ centered at the origin of each monomer and undergoes rotational diffusion. 
The translational diffusion coefficient $D$ is related to the temperature and the translational friction $\gamma$ via the Stokes-Einstein relation $D=k_{\rm B}T\gamma^{-1}$. Likewise, the rotational diffusion coefficient, $D_r=k_{\rm B}T\gamma_r^{-1}$, with $D_r = 3D\sigma^{-2}$. The  solvent induced Gaussian white-noise terms for both the translational $\pmb{\xi}$ and rotational $\pmb{\xi}_r$ motion are characterized by $\langle \pmb{\xi}(t)\rangle = 0$ and $\langle \xi_\alpha(t) \xi_\beta(t^\prime)\rangle = \delta_{\alpha\beta}\delta(t-t^\prime)$, where $\alpha,\beta \in \{x,y,z\}$. $\pmb{f}(t)$ includes the monomer-monomer interactions and the harmonic forces holding the polymer together. Harmonic bonds of the form $U_b=\frac{1}{2}k(r_{i,i+1}-\sigma)^2$ ensure chain connectivity. Here $r_{i,i+1}$ is the distance between consecutive monomers along the chain, and $k$ is the spring constant.
The monomer-monomer interaction forces are described by a Lennard-Jones (LJ) isotropic potential of the form
\begin{equation}
U(r_{ij})=4\varepsilon\left[ \left( \frac{\sigma}{r_{ij}}\right)^{12} - \left(\frac{\sigma}{r_{ij}}\right)^{6}\right ].
\label{pot}
\end{equation}
The interaction extends up to a cut-off distance of $2.5\sigma$, and is there to  impose self-avoidance between the monomers (the repulsive part) while simultaneously mimicking the effective attraction between the monomers due to the presence of a bad solvent. In our simulations $\sigma$ and $k_{\rm B}T_0$ are used as the units of length and energy scales of the system, while $\tau=\sigma^2D^{-1}$ is our unit of time. To quantify the strength of the active forces it is useful to introduce the dimensionless P\'eclet number defined as $Pe=v_p\sigma/D$. In this study, we consider a set of interaction strengths in the range $\varepsilon \in (1,10)k_BT_0$, and $k=500k_{\rm{B}}T_0/\sigma^2$ is set to be large enough to ensure polymer connectivity while simultaneously minimizing bond stretching that could arise from the action of the active forces. The equations of motion of the monomers were integrated using Brownian dynamics with a time step $\Delta t \in (10^{-4},2.5\times10^{-5})$ depending on the temperature and the active velocity considered, and all the simulations were iterated from $5\times10^8$ to $10^9$ steps.
 
\begin{figure}[h!]
\centering
\includegraphics[width=0.5\textwidth]{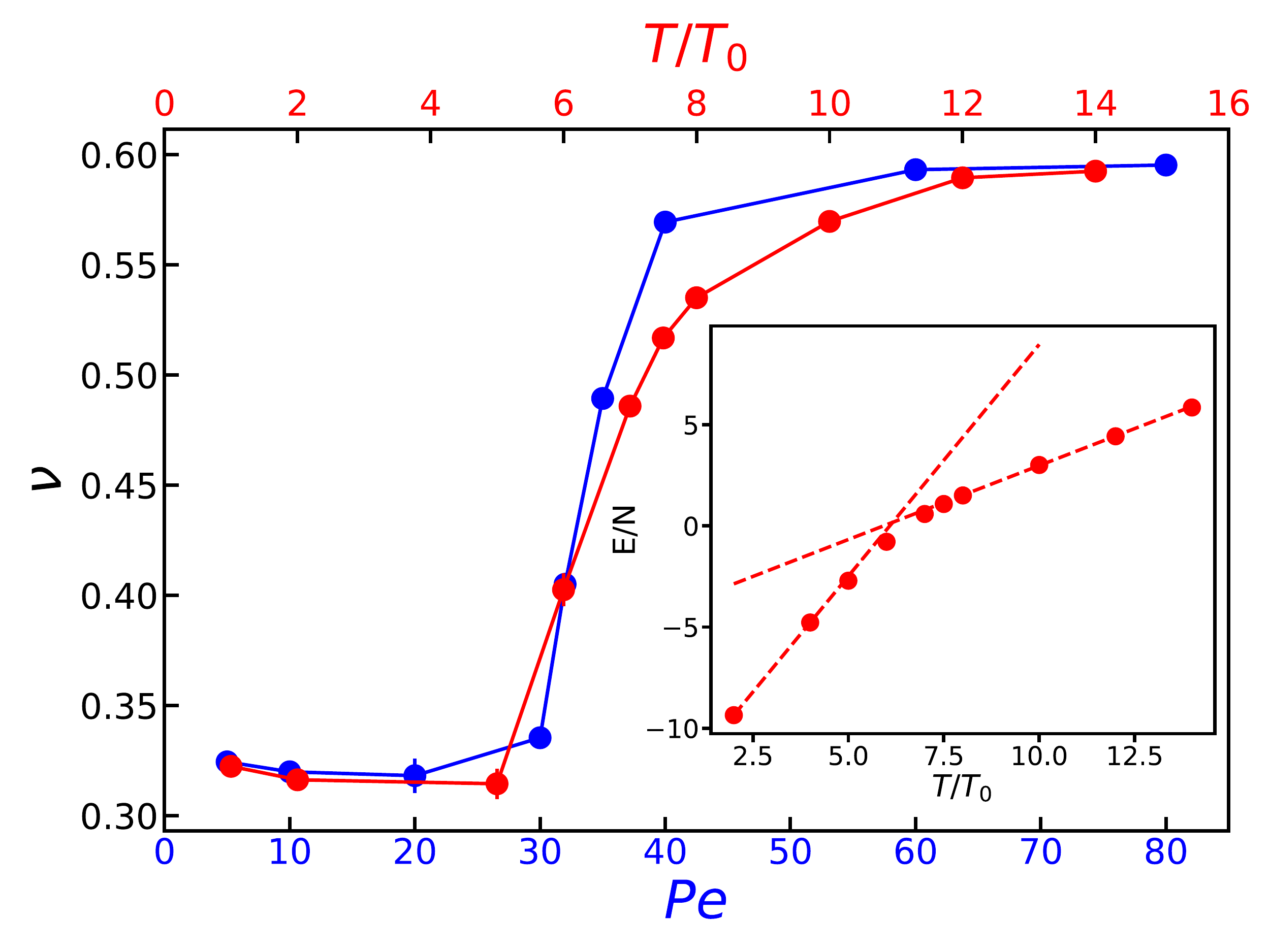}
 \caption{Scaling exponent $\nu$ for the activity-induced and temperature-induced coil-globule transition of a self-avoiding polymer with monomer-monomer interaction strength $\varepsilon=2.5k_BT_0$.
 The bottom horizontal axis refers to the data of an active polymer going through the transition as a result of activity fluctuations expressed in terms of the P\'eclet number $Pe$, whereas the top horizontal axis refers to the temperature-induced transition for a passive polymer resulting from a temperature change $T/T_0$. The inset shows the  energy of the polymer per number of monomers as a function of temperature for the passive polymer. In this case $E/N$ includes both the interaction energy between the monomers and the bond-stretching energy of the chain.}
 \label{fig1}
\end{figure}
\section{Results and Discussion}
We started our numerical simulations by characterizing the $\Theta$-point for our polymer model.
This is done by fixing the interaction energy to $\varepsilon=2.5k_BT_0$ and analyze the statistical properties of the polymer at different temperatures. To determine the scaling exponent of the polymer with its size under different system conditions, we used a combination of the scaling of the radius of gyration and the static structure factor defined as
\begin{equation}
    R_g=\frac{1}{N}\left\langle\sum_{i=1}^N\left(r_i-R_{cm}\right)^2\right\rangle; \quad
    S(q)=\frac{1}{N}\left\langle\sum_{j=1}^N\sum_{k=1}^N e^{-i\pmb{q}\cdot (\pmb{r}_j-\pmb{r}_k)} \right\rangle,
\end{equation}
where $R_{cm}$ is the center of mass of the polymer and $\pmb{q}$ is the wave vector.
In fact, for sufficiently large polymer chains, the structure factor satisfies the scaling relation
\begin{equation}
    S(q) \propto q^{-1/\nu},
\end{equation}
within the range $R_g^{-1}\ll q \ll \sigma^{-1}$. We considered polymers of length $N \in (32,512)$ for the radius of gyration scaling and N=512 for the static structure factor scaling. At low temperatures, we observe the expected  globular state of the polymer characterized by an exponent $\nu=1/3$. As the temperature is increased towards its $\Theta$ temperature and beyond, the size exponent increases up to a value around 0.58, the one expected for the coil state. Figure~\ref{fig1}, data in red, shows the size exponent $\nu$ as a function of temperature $T$ for the passive polymer. The inset in the same figure shows how the average energy per monomer of the chain changes with temperature~\cite{ciesla2007molecular}, and we find this to provide an accurate estimate of the $\Theta$ temperature of the polymers. 
Two distinct regions of different heat capacity can be distinguished from this plot of energy variation with temperature.
The $\Theta$ temperature is $k_BT_{\theta}=6.16$, corresponding to a ratio $\varepsilon/k_BT_{\theta}=0.406$.

Next, we considered the same analysis for an active   polymer and varied the magnitude of the activity, $Pe$, while keeping temperature and $\varepsilon$ constant. Figure~\ref{fig1}, in blue, shows how the scaling exponent changes with $Pe$.  Our results show a transition curve quite similar to that observed for the passive system.
When the strength of the active forces is much larger than of the monomer-monomer interactions, the polymer is well characterized as a flexible coiled structure with a size exponent $\nu\simeq 0.6$. Corresponding to a good-solvent condition for the active polymer. As the activity is lowered, a sudden transition to the globular, compact conformation, with a size exponent $\nu\simeq 1/3$ occurs.

\begin{figure}[h!]
\centering
\includegraphics[width=0.48\textwidth]{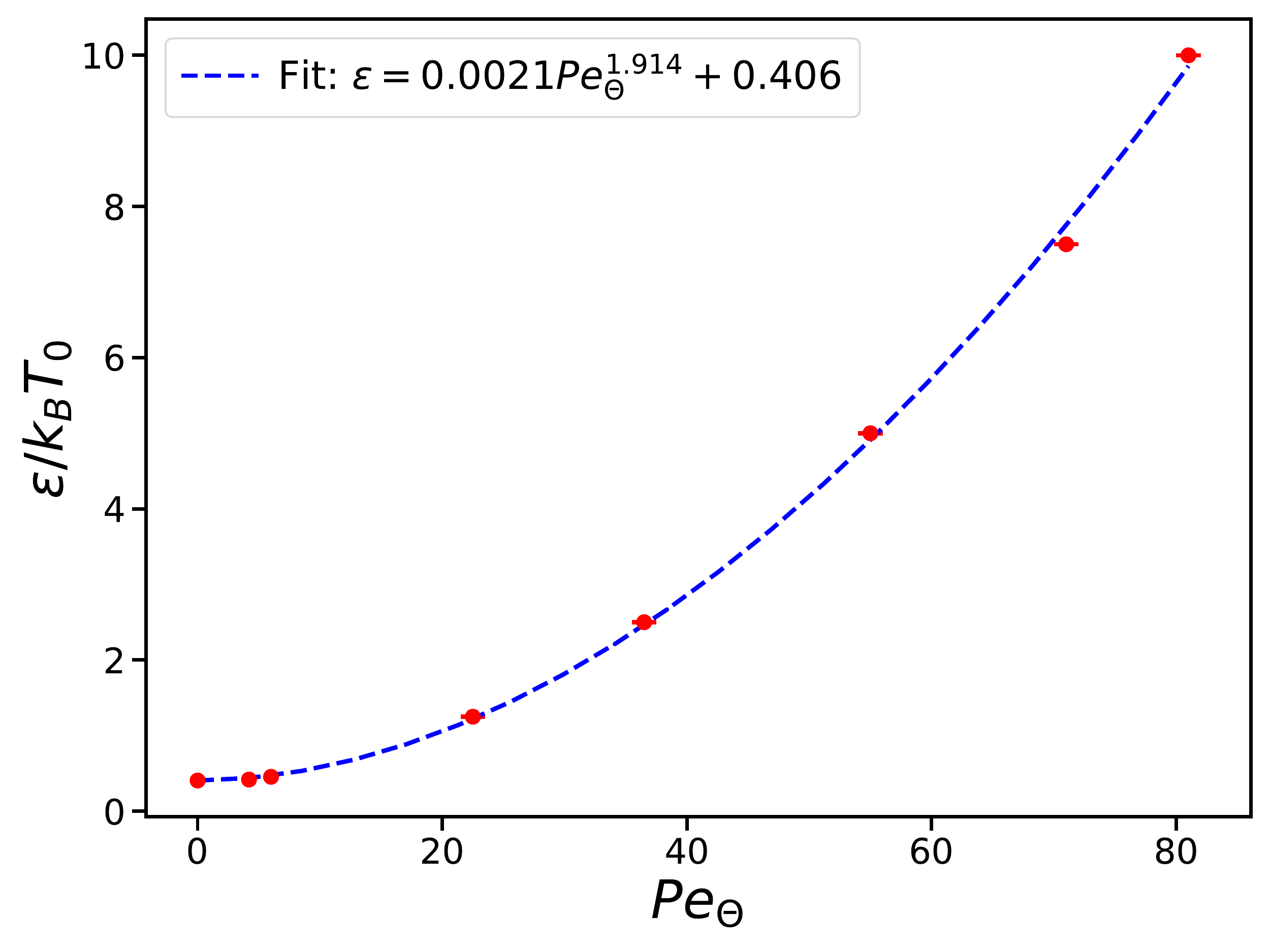}
 \caption{Plot showing the relation between the $\varepsilon$ and corresponding activity $Pe_\Theta$ at the $\Theta$-point of the activity-induced coil-globule transition of a polymer. The blue dashed line is a fit to the numerical data with a quadratic dependence.}
 \label{fig2}
\end{figure}

To further characterize this transition, we recomputed the transition curves for different values of the interaction strength $\varepsilon$ and sampled over various activities, and extracted the corresponding P\'eclet numbers, $Pe_{\Theta}$ close to the  $\Theta$ point of the polymers where $\nu=1/2$.
In Fig.~\ref{fig2} we show $Pe_{\Theta}$ for the different values of the interparticle attractive strength $\varepsilon$. The data are fitted
using a simple power law of the form 
$\varepsilon/(k_{\rm B}T_0)=0.406+b\,Pe_{\Theta}^{\delta}$.
We find that the exponent is compatible with a simple quadratic dependence $\delta=1.91(5) $ between the two parameters.

More insight into this relation can be established using simple theoretical considerations for the coil-globule transition. 
A very well known result from the Flory-Huggins\cite{FloryBook} theory for passive polymers is that the coil globule transition occurs when the second virial coefficient is equal to zero. This corresponds to setting the Flory-Huggins parameter, $\chi=1/2$, where $\chi$ is defined, in our case, via the relation 
$\int_\sigma U(r)d^3r \equiv (-4/3\,\pi\sigma^3)  k_{\rm B}T (2\chi)$. For our model, the $\Theta$ temperature is then obtained by setting $\varepsilon/(k_{\rm B}T_{\theta})\simeq 0.41$, which is very close to our numerical result for the passive case.

It was recently shown~\cite{kaiser_how_2015} that the role of activity in the statistical properties of an ideal fully flexible polymer can be taken into account by simply rescaling the temperature $T\rightarrow T(1+\frac{2}{9}Pe^2)$.
It is therefore tempting to extend this mapping  to our problem by also rescaling the  temperature  embedded in the definition of $\chi$ by such a factor. If we do that, then the condition $\chi=1/2$ leads to the relation 
\begin{equation}
\frac{\varepsilon}{k_{\rm B}T_0}=\frac{4\pi}{3A_0}(1+\frac{2}{9}Pe^2)\simeq 0.41+0.092 Pe^2\,,
\end{equation}
where in the last step we used $A_0\simeq 10.1$ to match the equation with our model. Here $A_0$ is the three-dimensional integral of our interaction potential in Eq~\ref{pot}  evaluated from $\sigma$ to $2.5\sigma$.
Alternatively, one could try to include the role of active fluctuations using 
the active energy scale $k_sT_s=\gamma v^2/(6D_r)$ introduced in~\cite{swimPressure}. 
In this case we have $k_{\rm B}T\rightarrow k_{\rm B}T +k_sT_s$. Using this approximation, and specifying the different parameters to our model, we find
\begin{equation}
\frac{\varepsilon}{k_{\rm B}T_0}=\frac{4\pi}{3A_0}(1+\frac{1}{18}Pe^2)\simeq 0.41+0.023 Pe^2\,,
\end{equation}
Although both of these simple temperature mappings appropriately capture the quadratic dependence of the $\Theta$ point on the strength of the active force, the pre-factor is almost off by a factor of fifty in the first case, and by a factor of ten in the second case, suggesting a  more complex interplay between active and thermodynamic forces.

\begin{figure}[h!]
\centering
\includegraphics[width=0.5\textwidth]{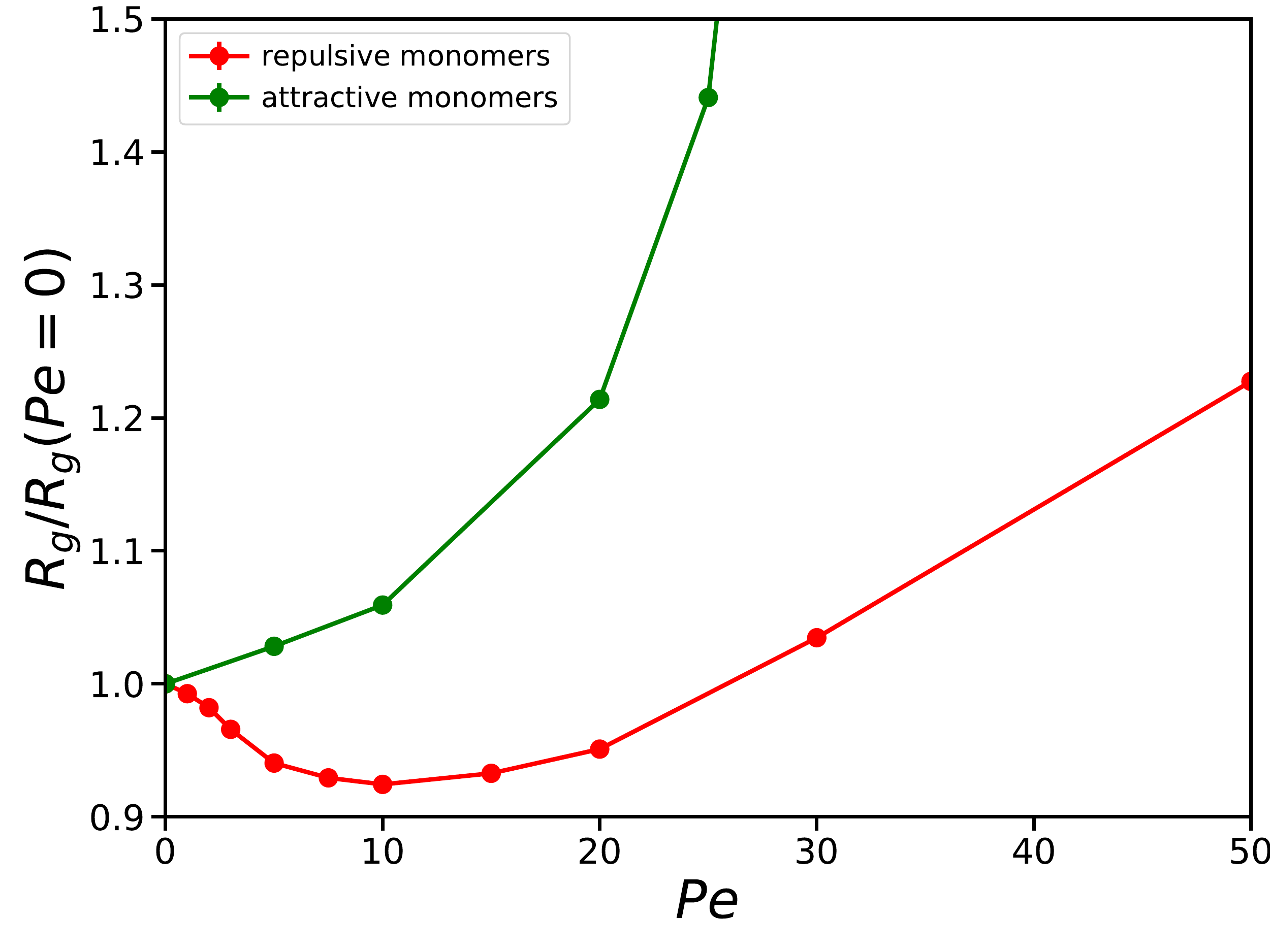}
 \caption{Radius of gyration $R_g$ for a flexible self-avoiding polymer (in red) and an attractive polymer (in green) as a function of the strength of the active forces  $Pe$. The data for the self-avoiding polymer are taken with a polymer of $N=256$ monomers in a good solvent  obtained by setting the cutoff of our interaction to $2^{1/6}\sigma$. The data for the attractive polymer contains $N=512$ monomer and an interaction strength set to $\varepsilon/k_{\rm B}T_0=2.5$ with cutoff at $2.5\sigma$. Both  data sets are normalized by the corresponding values of $R_g$ at $Pe=0$.}
 \label{rg_v}
\end{figure}

This mismatch should not come as a surprise. For instance, its already known, as we discussed earlier, that even for a simpler system, an active self-avoiding flexible polymer, the radius of gyration has a non-monotonic trend with the strength of the active forces, a behavior that cannot be understood with a simple temperature mapping of the activity~\cite{kaiser_active_2015,harder_activity-induced_2014,eisenstecken2017conformational,anand2020conformation}.

To make this more specific, we show in Fig.~\ref{rg_v} how the radius of gyration of a  self-avoiding polymer in a good solvent, obtained by setting the interaction cutoff in our model to $2^{1/6}\sigma$, depends on $Pe$ for a polymer of $N=256$ monomers. We observe a minimum of the radius of gyration for $Pe\simeq 10$, and importantly it shows how for $Pe<30$ the value of $R_g$ is smaller than that of its parent passive polymer. This result illustrates that for moderate activities, the self-avoiding polymer is always more compact than a passive polymer, and seems to suggest that adding an explicit attractive monomer-monomer interaction should lead to a collapse of the polymer for smaller values of $\varepsilon$ when compared to its  passive polymer counterpart. Our numerical results clearly show that this is not the case for any value of $Pe$. In the same figure, we also show the radius of gyration of an active polymer with attractive monomer-monomer interactions (initially in the collapsed state when $Pe=0$)  as a function of the Peclet number. In this case we observe a monotonic increase of  $R_g$ with $Pe$, suggesting that in this case activity always acts as an effective repulsion.
 
Although as discussed above a straightforward mapping does not seem obvious, it is nevertheless reasonable
to make an  analogy between our active polymer and a system of active  spheres. 
For an active suspension of hard spheres, it is well known that under the appropriate conditions an effective active negative pressure among the particles~\cite{swimPressure} can develop. This negative pressure, whose role becomes dominant for sufficiently large densities and  activities, is ultimately responsible for the motility induced phase separation (MIPS) observed in this system (see for instance \cite{MIPScates,swimPressure}).

The connectivity of the chain clearly makes our system different than an unconstrained suspension of active particles. In fact, for instance, we never observe a collapse of the chain in the absence of attractive forces (MIPS). However, we suspect a negative pressure-like term may also be present in this case, may be responsible at moderate activities for the size reduction  of the active self-avoiding polymer, and  it partially counteracts the enhanced active fluctuations that we mapped into an effective temperature. The net result should be a weakening of the $\Theta-Pe$ dependence as observed in our numerical calculations.
Interestingly, unlike the case of a self-avoiding polymer in a good solvent in two-dimensions, the ratio $R_g/R_g(Pe=0)$ is not always smaller than one for all P\'eclet numbers as observed in~\cite{kaiser_how_2015}, indicating that this negative active pressure must be relatively larger when the polymer is confined to fluctuate in a two dimensional plane. 

\section{Conclusions}
In this paper we considered the behavior of an active polymer subject to effective inter-particles interactions mimicking the properties of a good and bad solvent.
This study focused on the $\Theta$ temperature of an active polymer, and established its dependence on the strength of the active forces.

We found that a simple rescaling of the temperature provides a correct power law dependence (quadratic) between the ratio $\varepsilon/(k_{\rm B}T_0)$ and the active speed of the particles $Pe_{\Theta}$, however, the prefactor is at best one order of magnitude smaller than expected. Using a mean-field theory approach appropriate for passive polymers, we suggest that an effective negative pressure term, well established for suspension of active hard spherical particles, also exists for active polymers, has a quadratic dependence on $Pe$, and effectively increases the strength of the monomer-monomer interactions leading to a weaker  $\Theta-Pe$ dependence than expected from a simple temperature rescaling. 

It should be stressed that it is quite remarkable that a simple re-scaling of the temperature is capable of capturing the most essential scaling features of an active polymer even at its $\Theta$ point. Although we have not considered the effect of hydrodynamic interactions in this study, as our goal was to first understand within the framework of dry active matter the interplay between active and thermal fluctuations, it would be interesting to perform such a study and evaluate whether significant differences in the scaling laws may appear. Efforts in this direction are underway.

\section*{Conflicts of interest}
There are no conflicts to declare.

\section*{Acknowledgements}
A.C. acknowledges financial supported from the National Science Foundation under Grant No. DMR-1703873.



\balance


\bibliography{rsc} 
\bibliographystyle{rsc} 

\end{document}